\documentstyle[12pt]{article}

\input epsf

\def\beq{\begin{equation}}
\def\eeq{\end{equation}}

\def\al{\alpha}
\def\bt{\beta}
\def\Ga{\Gamma}
\def\de{\delta}
\def\De{\Delta}

\def\lam{\lambda}

\def\l{\left (}
\def\r{\right )}
\def\fr{\frac}
\def\la{\label}

\def\ov{\overline}
\def\ran{\rangle}
\def\lan{\langle}
\def\ti{\tilde}

\begin{document}

\begin{titlepage}

\begin{center}
{\Large \bf  Naturally Light Neutrinos and Unification \\
in Theories with Low Scale Quantum Gravity
}
\end{center}
\vspace{0.5cm}
\begin{center}
{\large Zurab Tavartkiladze}\footnote{%
E-mail address: z\_tavart@osgf.ge
}  
\vspace{0.5cm}

{\em Institute of Physics, Georgian Academy of Sciences,
380077 Tbilisi, Georgia}\\
\end{center}

\vspace{1.0cm}

\begin{abstract}

Within low scale theories traditional see-saw and scalar triplet
mechanisms, for neutrino mass suppression, do not work out anymore and
for realistic model building some new ideas are needed. 
In this paper we suggest mechanism, different from existing ones, which
provides natural suppression of the neutrino masses. The mechanism is
realized through extended scalars of $4$, $5$ or $6$ dimensional $SU(2)_L$
multiplets. Scenario, with fundamental mass scale $M_f$ in a 
$\sim 10^{3}$~TeV range, requires $4$-plets guaranteeing neutrino masses 
$\stackrel{<}{_{\sim }}1 $~eV. For theories with
$M_f={\rm few}\cdot 10$~TeV $5$-plets should be involved, while in
scenarios with $M_f=$~few TeV, $6$-plets could be efficient.

The considered mechanism could be successfully applied also for
supersymmetric theories, building scenarios with various values of
low $M_f$.

Within considered models we also address the question of gauge coupling
unification. For low scale unification, existence of compact
extra dimensions turns out to be crucial. Due to additional scalar
multiplets,
some new examples of unification are found for both - non SUSY and SUSY
cases. Within non SUSY scenarios introduced $SU(2)_L$ scalars take
advantage and are important for successful unification.

\end{abstract}
\end{titlepage}

\section{Introduction}

Theories with extra spacetime dimensions have attracted great attention
{}for last years. Main phenomenological motivation, {}for considering such
type of scenarios,
was the new possibilities of resolution of gauge hierarchy problem
\cite{tevgr1}-\cite{gogb}.
It was observed \cite{tevgr1} that due to appropriately large
extra dimensions, it is possible to lower fundamental scale $M_f$ even
down to few TeV, while the observed weakness of gravity
could be explained through large volume of extra space. In fact, for extra
spacelike dimension's number $\de=2$, four dimensional Planck mass
$M_{\rm Pl}$ has value $\sim 10^{19}$~GeV if size $R$ of extra compact
dimensions is in a range $\stackrel{<}{_\sim }1$~mm (distance at which
the behavior of gravity is still unknown and is extensively investigated
in upgoing experiments \cite{gravexp}).

An alternative approach has been suggested in Refs. \cite{RS, gogb}.
In these scenarios, although the fundamental scale can be close to 
$M_{\rm Pl}$, the required hierarchy is obtained on a `visible' brane
through non-factorizable geometry. Latter solution emerges from higher
dimensional gravity with one \cite{RS, gogb} or more \cite{mult}
extra dimensions.

In both type, of above mentioned scenarios, it is assumed that all Standard
Model (SM) fields are confined to a $3$-brane (identified with our
Universe) in extra dimensions. The idea, that we
live on a brane/topological defect embedded in a higher dimensional space,
goes back to \cite{univ}.

Despite great success in solving the gauge hierarchy problem,
within this type of theories, various
problems arise and numerous issues require to be reconsidered from a new
viewpoint.
This cast an intriguing challenge to theoreticians. Amongst
raised issues, the actual task is to understand how to suppress 
neutrino masses in
a needed level.  Due
to low fundamental scale, 
the well known see-saw \cite{seesaw} and scalar triplet \cite{triplet}
mechanisms do not lead to the sufficient suppressions. 
Of course, by requiring conservation of lepton number, it is possible to
restrict operators responsible for neutrino masses. On the other hand, 
latest atmospheric \cite{atm} and solar \cite{sol} neutrino data have
increased confidence in the neutrino oscillations. So, the purpose is to
generate neutrino masses with desirable magnitudes.
In \cite{bulknu1} there were suggested mechanisms, which have
extra dimensional nature and successfully resolve this problem.
One way is to couple bulk right handed neutrinos with left handed ones. In
this case suppression occurs in Dirac Yukawa couplings through the large
volume factor \cite{bulknu1, bulknu2}. Needed suppression also can be
achieved if lepton number
violation takes place on a distant brane \cite{bulknu1}.
Different approach was presented in \cite{neutsup}, where for suppression
of neutrino masses, together with right handed neutrinos was introduced
additional scalar doublet with a sufficiently tiny VEV.

In this paper we suggest mechanism, which provides generation of 
suppressed neutrino masses. The mechanism do not has extra
dimensional nature. Suppression occurs due to group-theoretical reason.
Namely,
by introducing extended charged $ SU(2)_{L} $ scalar multiplets
in $4$, $5$ or $6$ representations, neutrinos gain masses of a needed
value.
$4$ dimensional plets are efficient if fundamental
scale lies in a range  $\sim 10^{3}$~TeV. $5$-plets
are motivated {}for $M_f={\rm few}\cdot 10$~TeV, while for $M_f=$~few TeV
$6$-plets should be involved.
Suggested mechanism can be successfully applied also for low scale SUSY
theories.

We also address question of low scale unification which,
within low scale theories, has different insight. Due to presence of extra
compact dimensions, $SU(3)_c$, $SU(2)_L$, $U(1)_Y$ gauge couplings will
have power low running \cite{ven} above the scale $1/R$. This give
possibility to obtain low scale unification
\cite{uni1}-\cite{uni3}. 
For our scenarios, if masses of introduced scalars lie above the
GUT scale, then they do not alter renormalization and status of
unification would be same as for cases of Refs. \cite{uni1}-\cite{uni3}.
However, if masses
of $4$, $5$ or $6$ plets (depending on a considered scenario)
are below the GUT scale, the situation is changed.
This open up new possibilities of unification  for
non SUSY and SUSY scenarios as well. Within non SUSY models introduced
$SU(2)_L$ scalars 
are crucial for successful unification.

\section{Suppressed neutrino masses }

For generating adequately suppressed neutrino masses, within low scale
theories, we
will introduce extended $SU(2)_L$ scalars. 

Introduce scalar $\Phi $ which under $SU(2)_L\times U(1)_Y$ transform as
$(4,~-3)$, where $U(1)_Y$ charge is measured in the units of charge of
lepton doublet $l$.
Our studies, of generation suppressed neutrino masses, do
not related
with existence of extra dimensions. We assume that $\Phi $ and all
standard model particles are localized on a $3$-brane.
In order to avoid too large Majorana neutrino masses, somehow we have to
forbid 
$(lh^{+})^2/M_f$ type operators ($h$ is SM Higgs
doublet). This can be achieved through some symmetries
\footnote{In fact, in the Yukawa
sector, responsible for generation of charged fermion masses, lepton
number is accidentally conserved. Assuming that, this conservation has
some fundamental origin, one can extend $L$ conservation also to the
appropriate $d=5$ operators.}.
For simplicity we will assume that lepton number ($L$)
is conserved in the fermion sector (as were done in
\cite{bulknu1}-\cite{neutsup}) and prescribe
to $\Phi $ lepton number $-2$. So, the Yukawa sector possesses $U(1)_L$
symmetry:

\beq
{\cal L}_{\nu }=\fr{\hat{\lam}_{\nu }}{M_f}ll \Phi h+h.c.~,
\la{4nulag}
\eeq
where $\hat{\lam}_{\nu }$ is dimensionless matrix in a family space. As we
see, in this case, neutrino masses get additional suppression  
$\lan h\ran /M_f $ (in comparison with scenario with scalar triplets
\cite{triplet}). But
for low $M_f$ more suppression is needed. Namely, $\Phi $ should develop
appropriately tiny VEV along its neutral component. This is naturally
insured
through the scalar potential, with relevant couplings. Most general 
$SU(3)_c\times SU(2)_L\times U(1)_Y$ invariant renormalizable potential is 

$$
{\cal V}(h, \Phi )=\fr{\lam_h}{4}\l h^{+}h-m^2\r^2+
\fr{\lam_{\Phi }}{4}\l \Phi^{+}\Phi +M^2\r^2+
$$ 
\beq
\fr{\lam_1}{2}\l \Phi^{+}\Phi \r \l h^{+}h \r + 
\fr{\lam_2}{2}\l \Phi^{+}h \r \l\Phi h^{+} \r - 
\fr{\lam }{2}\l \Phi h^{3}+\Phi^{+}(h^{+})^3\r ~,
\la{4pot} 
\eeq
where all parameters are assumed to be positive. $m$ is Higgs doublet mass
of the order of $\sim 100$~GeV, while $M$ is mass of $\Phi $ 
field. 
Last term in (\ref{4pot})
mildly violates $U(1)_L$. It involves highest power of $h$ and therefore,
between last three intersecting quartic terms, will be most
suppressed. This is in a good accordance with a so-called naturalness
issue \cite{nc}.
%
{}For $\lam >0$, 
system will have global minimum with non zero $\lan \Phi \ran $. 
%
The extremum conditions for (\ref{4pot}) will be: 

$$
\lam_h(v^2-m^2)+(\lam _1+\lam _2)V^2-3\lam Vv=0~,
$$
 \beq
\lam_{\Phi }(V^{2}+M^{2})V+(\lam_1+\lam_2)Vv^2-\lam v^3~=0~.
\la{min} 
\eeq
{}For all positive parameters in (\ref{4pot}) and {}for

\beq
\lam_{\Phi }M^{2}\gg (\lam_1+\lam_2)m^2~, 
\la{cond} 
\eeq
one can easily obtain

\beq
v=m+{\cal O}\l m^3/M^2\r ~, ~~~~ 
V=\fr{\lam }{\lam_{\Phi }}\l \fr{v}{M}\r^2v+{\cal O}\l m^5/M^4\r ~. 
\la{4sol} 
\eeq
Note, that although the mass of $\Phi $ is much larger than $v$, the
hierarchy is not destabilized, because $\Phi$'s VEV in (\ref{4sol}) is
tiny and quartic terms in (\ref{4pot}) practically do not affect $v$.
For $h$'s potential [first term in (\ref{4pot})] we have used the simplest
possible expression. The potential's form for SM doublet is not crucial,
because VEV of $\Phi $ will have same magnitude as in (\ref{4sol}).
Important is to achieve desirable electroweak (EW) symmetry
breaking. Since this issue
is beyond the scope of this paper, we will assume that one of the
mechanisms \cite{EWbr}, \cite{EWbr1}, providing EW symmetry breaking
(EWSB), for
low scale theories, is applied. The natural hierarchy between EW and
fundamental scales can be achieved if EWSB occurs
dynamically \cite{EWbr}, wile SUSY theories (which we consider
below) guarantee stability of the scales [condition in (\ref{cond})].

Using (\ref{4sol}) in (\ref{4nulag}), for neutrino masses we will have

\beq
\hat{m}_{\nu }=\hat{\lam }_{\nu }\fr{V}{M_f}v\simeq 
\fr{\lam \hat{\lam }_{\nu }}{\lam_{\Phi }}
\l \fr{v}{M}\r^3\fr{M}{M_f} v~, 
\la{numas4} 
\eeq
and desirable value $\hat{m}_{\nu }=(1-4\cdot 10^{-2})$~eV is obtained 
for $M\simeq M_f=(1-3)\cdot 10^{3}$~TeV with $v=174$~GeV, 
$ \lam \hat{\lam }_{\nu }/\lam _{\Phi }\sim 1 $. This scale for neutrino
masses is natural for atmospheric anomaly \cite{atm} if three family
neutrinos are either degenerate in mass or hierarchical,
respectively. Smaller scale, relevant for solar neutrinos \cite{sol}, can
be obtained through suppressing the appropriate entries in 
$\hat{\lam }_{\nu }$. Latter can be naturally realized through the flavor
symmetries.

If we wish to build scenario with lower fundamental scale, higher
$SU(2)_L$ representations must be introduced. Namely, if now $\Phi $ is
$5$-plet of $SU(2)_L$ with $U(1)_Y$ charge $-4$, then Yukawa couplings,
responsible for neutrino masses will be

\beq
{\cal L}_{\nu }=\fr{\hat{\lam }_{\nu }}{M_f^{2}}ll \Phi h^{2}+h.c.~,   
\la{5nulag}
\eeq
and in potential (\ref{4pot})
last term will be replaced with
$-\fr {\lam' }{2M_f}\l \Phi h^{4}+\Phi ^{+}{h^{+}}^{4}\r $.
For this case it is easy to verify

\beq
v\simeq m~,~~~~~ 
V\simeq \fr{\lam '}{\lam_{\Phi }}\l \fr{v}{M}\r^3\fr{M}{M_f}v~.
\la{5sol} 
\eeq
Using (\ref{5nulag}) and  (\ref{5sol}), for neutrino masses we will have

\beq
\hat{m}_{\nu }=\hat{\lam }_{\nu }\fr{V}{M_f^2}v^2\simeq
\fr{\lam' \hat{\lam }_{\nu }}{\lam_{\Phi }}\l \fr{v}{M}\r^5 
\l \fr{M}{M_f}\r^3v~, 
\la{numas5} 
\eeq
which for $\hat{m}_{\nu }=(1-0.1)$~eV, 
$\lam' \hat{\lam }_{\nu }/\lam _{\Phi }\sim 1 $
require relatively low scales  $M\simeq M_f=(30-50)$~TeV.

Fundamental scale can be easily reduced even down to few TeV, if 
$\Phi $ belongs to $(6,~-5)$ representation of $SU(2)_L\times U(1)_Y$.
Then instead the last term in (\ref{4pot}) we will have 
$-\fr {\lam'' }{2M_f^2}\l \Phi h^{5}+\Phi ^{+}{h^{+}}^{5}\r $
and relevant Yukawa couplings will be
$\fr{\hat{\lam }_{\nu }}{M_f^{3}}ll \Phi h^{3}$. By simple analyses one
can
easily obtain that in this case

\beq
\hat{m}_{\nu }\simeq
\fr{\lam'' \hat{\lam }_{\nu }}{\lam_{\Phi }}\l \fr{v}{M}\r^7
\l \fr{M}{M_f}\r^5v~,
\la{numas6}
\eeq
and $(1-0.1)$~eV neutrino masses 
(for $\lam'' \hat{\lam }_{\nu }/\lam _{\Phi }\sim 1$) is generated for
$M\simeq M_f=(7-10)$~TeV.

\vspace{0.2cm}

These scenarios can be successfully extended to the low scale
supersymmetric
theories. In SUSY versions, together with chiral superfield $\Phi $
(which denote $4$, $5$ or $6$-plets) must be introduced conjugate 
$\ov{\Phi } $ supermultiplet. The relevant superpotential will be

\beq
W_{\Phi }=M\ov{\Phi } \Phi-\fr{1}{M_f^{1+n}}
\l \lam_{\Phi d}\Phi h_d^{3+n}+
\lam_{\Phi u}\ov{\Phi } h_u^{3+n}\r~,
\la{sup}
\eeq
where $n=0, 1, 2$ for scenarios with $\Phi+\ov{\Phi }$
in $4$, $5$ and $6$ representations of $SU(2)_L$ respectively.
$h_u$, $h_d$ denote doublet-untidoublet pair of MSSM and 
$\lam_{\Phi d}$, $\lam_{\Phi u}$ are positive dimensionless coupling
constants. Yukawa superpotential, responsible for neutrino masses, will be

\beq
W_{\nu }=\fr{\hat{\lam}_{\nu }}{M_f^{n+1}}ll\Phi h_d^{n+1}~.
\la{yuksup}
\eeq
In unbroken SUSY and EW symmetry limit 
$\lan h_u \ran=\lan h_d\ran =0$, and from (\ref{sup}) follows
also $\lan \Phi \ran =\lan \ov{\Phi }\ran =0$.
After that SUSY and EW symmetry breaking take place\footnote{Still, it is 
assumed
that one of the mechanisms, for SUSY and EW symmetry breaking, is applied
(see \cite{susybr} and \cite{EWbr}, \cite{EWbr1} respectively).}, 
non zero $\lan h_u \ran,~\lan
h_d\ran $
are generated and from (\ref{sup}) one can easily verify
$\lan \Phi \ran \simeq \lam_{\Phi u}\lan h_u\ran^{n+3}/(MM_f^{n+1})$.
Using this and also (\ref{yuksup}), for neutrino masses we will get

\beq
\hat{m}_{\nu }=\hat{\lam }_{\nu }\lam_{\Phi u}
\l \fr{v}{M_f}\r^{2n+2}
\fr{v^2}{M}\sin^{n+3}\bt \cos^{n+1}\bt~,
\la{susynumas}
\eeq
where we have used $\lan h_u\ran=v\sin \bt $,
$\lan h_d\ran=v\cos \bt $. As we see, within SUSY scenarios expressions
for neutrino masses are slightly modified [compare with (\ref{numas4}),
(\ref{numas5}), (\ref{numas6})]. However, needed suppressions are still
guaranteed. In particular, for 
$\hat{\lam }_{\nu }\lam_{\Phi u}\stackrel{<}{_\sim }1$, $v=174$~GeV
and $\tan \bt \simeq 1$, neutrino masses 
$m_{\nu }\stackrel{<}{_\sim } (1-0.1)$~eV are obtained within various
scenarios:

\begin{equation}
M\simeq M_f=\left\{ \begin{array}{lll}
(0.6-1.3)\cdot 10^3~{\rm TeV}; & n=0,~ {\rm case~ with
~4-plets} \\
(20-30)~{\rm TeV}; & n=1,~ {\rm case~ with~ 5-plets} \\
(4.7-6.5)~{\rm TeV}; & n=2,~ {\rm case~ with~ 6-plets}
\end{array} 
\right.~.
\la{susyrange}   
\end{equation}
Larger values of $\tan \bt $ would give stronger suppression of neutrino
masses in (\ref{susynumas}), giving possibility to reduce mass scales in
(\ref{susyrange}) by few factors.

Obtaining ranges for scales, in (\ref{numas4}),
(\ref{numas5}), (\ref{numas6}) and (\ref{susyrange}), we have assumed
$M\simeq M_f$
(and $\tan \bt \simeq 1$ for SUSY cases). Obviously,
it is possible to have $M_f$ by few factors larger than the value of $M$.
This will slightly modify the ranges for mass scales. 
Important point is, that
mechanisms which we have suggested here, provide adequate suppressions of
neutrino masses and this suppressions occur through proper choice of
scalar $\Phi $ in appropriate $SU(2)_L\times U(1)_Y$ representation.

\section{Gauge coupling unification}

If extra spacelike dimensions exist, it is possible to obtain the
low scale unification of gauge coupling constants. This can take place
if scale $\mu _{0}= 1/R<M_{G}$.
Above the scale $\mu _{0}$
the heavy Kaluza-Klein (KK) states enter into the game and gauge
coupling runnings
become power low. The solution of one loop RGEs
have forms \cite{ven}-\cite{uni3}:

\beq
\al^{-1}_G=\al_a^{-1}- \fr{b_a}{2\pi }\ln \fr{M_G}{M_Z}+\De_a~,
\la{alps} 
\eeq
where $\al _{1,2,3}$ denote gauge couplings (on scale $M_{Z}$)
of $U(1)$, $SU(2)_{L}$ and $SU(3)_{c}$ respectively,
$b_{a}$ is standard b-factors (depending which theory we are
studying - non-SUSY or SUSY). In general, in $\De _{a}$ could
contribute two type of terms

\beq
\De_a=\De_a^{0}+\De_a^{KK}~, 
\la{sumde} 
\eeq
where $\De_a^0$ denote contribution of some additional states
with masses $M_i$ below the GUT scale, and have logarithmic energy
dependence

\beq
\De_a^{0}=-\fr{\ti b^i_{a}}{2\pi }\ln \fr{M_{G}}{M_i}~. 
\la{de0} 
\eeq
$\De _{a}^{KK}$ express contribution of KK states and
have power low energy dependence \cite{ven}

\beq
\De_a^{KK}=-\fr{\hat{b}^i_{a}}{2\pi }P_{\de }^{(\mu_i)}~,~~~~~~ 
P_{\de }^{(\mu_i)}=\fr{X_{\de }}{\de } \left [\l
\fr{M_G}{\mu_i}\r^{\de }-1\right ]-
\ln \fr{M_G}{\mu_i}~, 
\la{deKK} 
\eeq
where $X_{\de }=\pi ^{\de /2}/\Ga (1+\de /2)$, $\mu _i^2=M_i^2+\mu_0^2=
M_i^2+1/R^2$
(for SM and MSSM states $M_i=0$).
{}For simplicity we have assumed that all $\de $ compact extra spacelike
dimensions have equal radius. 
{}From (\ref{alps}), excluding $\al_G$ and $\ln (M_G/M_Z)$, for strong
coupling we find:

\beq
\al_s^{-1}=\fr{b_1-b_3}{b_1-b_2}(\al_2^{-1}+\De_2)-
\frac{b_2-b_3}{b_1-b_2}(\al_1^{-1}+\De_1)-\De_3~,
\la{alpsgen}
\eeq
and for given values of $\De_i$ we can estimate the value of
$\al_s(M_Z)$. Also, through (\ref{alps}) one can calculate the value
of GUT scale

\beq
\ln \fr{M_G}{M_Z}=\fr{2\pi }{b_1-b_2}
\l \al_1^{-1}-\al_2^{-1}+\De_1-\De_2 \r~,
\la{scales}
\eeq
and finally, the value of unified gauge coupling constant 

\beq
\al_G^{-1}=\fr{1}{b_1-b_2}\left [ b_1(\al_2^{-1}+\De_2)-
b_2(\al_1^{-1}+\De_1)\right ]~.
\la{alunif}
\eeq
Through analyzes  one has to make sure that $\al_G$ remains in a
perturbative regime.

In case with $\ti{b}^i_{a}=0$ (or $M_i\stackrel{>}{_{-}}M_{G}$)
and $\de =0$, $\De _{a}^{0}=\De _{a}^{KK}=0$ and the
status of unification is unchanged. In case, when there are no additional
states (e.g. $\De _{a}^{0}=0$) and we have only KK excitations
(either of SM or MSSM states),
the unification picture is not altered if $\hat{b}_{a}$ factors
satisfy condition:

\beq
\fr{\hat{b}_{a}-\hat{b}_{b}}{b_{a}-b_{b}}=const.~~~~
 ({\rm for}~ a\neq b)~.
\la{condb} 
\eeq
{}For SM $b_a=(41/10, -19/6, -7)$, and case with
$\De_a=0$ predicts $\al_s=0.071$, which is unacceptable
\cite{pd}. As was pointed
out in \cite{uni1}, the existence of extra dimensions open up
possibilities for improving this situation. Namely, if conditions in
(\ref{condb}) are mildly violated then one can attempt to get successful
low scale unification. This take place by introducing three real $SU(2)_L$
adjoint scalars only with KK excitations and no zero mode
wave functions\footnote{This is fully consistent with an orbifold
compactification scenarios. We do not go through latter issue here and
refer the reader to \cite{uni1}, where detailed discussions are
presented.}. 

In MSSM, in one loop approximation, we have successful picture of
unification and for its preserving conditions in (\ref{condb}) must be
satisfied (at least in a high accuracy). This can be reached by
introducing additional pairs of 
vectorlike chiral superfields \cite{uni2, uni3}.

{}For our scenarios, either with $4$, $5$ or $6$-plets, all above mentioned
cases of unification will be achieved if their masses $M$ are not below
the GUT scale. Successful unification will take place if the ideas of
\cite{uni1} and \cite{uni2, uni3} will be applied for non-SUSY and SUSY
cases
respectively. However, cases with $M<M_G$ will give different results and
we would like to study these examples here.

Let us start with non-SUSY case with $4$-plets. As it will turn out, these
states are crucial for unification. 
For $n_{\Phi }$ $4$-plets with
masses $M<M_G$ we have

\beq
\ti{b}_a=(\fr{9}{5},~ \fr{5}{3},~ 0)n_{\Phi }~, ~~~~~
\hat{b}_a=\hat{b}_a^{\rm SM}+(\fr{9}{5}, ~\fr{5}{3},~ 0)n_{\Phi }~,
\la{tilb4}
\eeq
where $\hat{b}_a^{\rm SM}=(1/10, -41/6, -21/2)+(8/3, 8/3, 8/3)\eta $
($\eta $ is number of chiral families with KK excitations).

Ignoring $\De^{KK}_a$, we will have  
$\al_s^{-1}=(\al_s^{-1})^0_{\rm SM}-
\fr{87}{109\pi}n_{\Phi}\ln \fr{M_G}{M}$, and for
$M_G/M\simeq 40$, $n_{\Phi }=6$, we obtain $\al_s=0.119$.
But without KK excitations there is no
power low
running and according to (\ref{scales}) no low scale unification is
obtained\footnote{See however \cite{nonextra}, where
examples of low
scale unification, without extra dimensions, were presented.}. 
As it turns out, unification take place for $\mu_0<M<M_G$. It means that
we have to include KK excitations of $\Phi (4)$ states starting only from
scale
$M$, while KK states of SM particles enter into the game from $\mu_0$
scale. For $\al_s$ we get:

\beq
\al_s^{-1}=(\al_s^{-1})_{\rm SM}^0-\fr{87}{109\pi }n_{\Phi }\ln
\fr{M_G}{M}-
\fr{87}{109\pi }n_{\Phi }P_{\de }^{(M)}-
\fr{1}{218\pi }P_{\de }^{(\mu_0)}~,
\la{alsKK4}
\eeq
where $P_{\de }^{(M)}$ and $P_{\de }^{(\mu_0)}$ denote functions
presented in (\ref{deKK}), calculated for appropriate scales. In
(\ref{alsKK4}) the $n_{\Phi }$ and ratios $M_G/M$, $M_G/\mu_0$ must be
chosen in such a way as to get desirable value for $\al_s$. At the same
time from (\ref{scales}) for $M_G$ we should get reasonable value (not too
larger than $M$) and also $\al_G$ in (\ref{alunif}) must be in a
perturbative
regime. Also, the values of $M$, $M_f$ should be such that neutrino
masses must be properly suppressed in (\ref{numas4}). 
We will assume that $M_G\simeq M_f$ and require 
$m_{\nu }\stackrel{<}{_\sim }1$~eV.
Taking into account all this,
from (\ref{alsKK4}), (\ref{scales}), (\ref{alunif}) it is easy to see that
successful unification
with $\al_s\simeq 0.119$ is obtained for $n_{\Phi }=2$ and various
values of extra dimensions and mass scales:

$$
\l \de ~, ~\fr{M_G}{\mu_0}~,~\fr{M_G}{M}~,~M_G\r=
\l 1,~~~9.78,~~~6.51,~~~10^{3.51}~{\rm TeV}\r ~,
$$
\beq
\l 2,~~~3.45,~~~2.83,~~~10^{3.26}~{\rm TeV}\r ~,~
\l 3,~~~2.36,~~~2.071,~~~10^{3.18}~{\rm TeV}\r ~,\cdots
\la{sol4}
\eeq 
The values of $\al_s$ and $M_G$ are $\eta $ independent, while $\al_G$
in (\ref{alunif}) depends on $\eta $. In this case for 
$0\stackrel{<}{_-}\eta \stackrel{<}{_-}3$ the $\al_G$ remains in a
perturbative regime $2\cdot 10^{-2}\al_G<4\cdot 10^{-2}$. Result of
numerical calculation for $\de=1$, $\eta=0$ is presented on Fig. 1, (a).

Similar discussions and analyses can be done for non-SUSY scenario with
$5$-plets. In this case

\beq
\al_s^{-1}=(\al_s^{-1})_{\rm SM}^0-\fr{325}{218\pi }n_{\Phi }\ln
\fr{M_G}{M}- 
\fr{325}{218\pi }n_{\Phi }P_{\de }^{(M)}-
\fr{1}{218\pi }P_{\de }^{(\mu_0)}~,
\la{alsKK5}
\eeq
and $\al_s\simeq 0.119$ is obtained for $n_{\Phi }=1$, with

$$
\l \de ~, ~\fr{M_G}{\mu_0}~,~\fr{M_G}{M}~,~M_G\r=
\l 1,~~~11.55,~~~6.9,~~~10^{1.82}~{\rm TeV}\r ~,
$$
\beq
\l 2,~~~3.735,~~~2.92,~~~10^{1.67}~{\rm TeV}\r ~,~
\l 3,~~~2.486~,~2.12,~~~10^{1.61}~{\rm TeV}\r ~,\cdots
\la{sol5}
\eeq 
Also in this case for $0\stackrel{<}{_-}\eta \stackrel{<}{_-}3$, the
$\al_G$ remains in perturbative regime 
($\stackrel{<}{_\sim }4\cdot 10^{-2}$).
Unification picture for this scenario, for $\de=1$, $\eta=0$, is
illustrated on Fig. 1, (b).

As far, the scenario with $6$-plets is concerned, unification 
near few TeV energies is obtained (with $\al_s\simeq 0.119$) for 
$n_{\Phi }=1$ and

$$
\l \de ~, ~\fr{M_G}{\mu_0}~,~\fr{M_G}{M}~,~M_G\r=
\l 1,~~~12.25,~~~4.55,~~~10.7~{\rm TeV}\r ~,
$$
\beq
\l 2,~~~3.837,~~~2.35,~~~8.69~{\rm TeV}\r ~,~
\l 3,~~~2.486~,~2.12,~~~8.11~{\rm TeV}\r ~,\cdots
\la{sol6}
\eeq
Fig. 1, (c) corresponds to this case with $\de =1$, $\eta =0$.

Let us now turn to the SUSY cases. In higher dimensional theories all
introduced states must be embedded in $N=2$ supermultiplets. We will
assume that MSSM doublet-untidoublet form one $N=2$ supermultiplet 
$(h_u,~h_d)$. And also each pair of $\Phi+\ov{\Phi }$
form one $N=2$ supermultiplet $(\Phi, ~\ov \Phi )$.
Since one loop value of $\al_s$ in MSSM is $0.117$, the new contributions
within our scenarios should not be large. As it turn out, for successful
unification, either with $4$, $5$ or $6$ supermultiplets, we have to
introduce
one additional state in the $SU(3)_c$ adjoint representation only with KK
excitations and without zero mode wave function. With this, for 
$n_{\Phi}$ pairs of $4+\bar 4$ we have:

\beq
\al_s^{-1}=(\al_s^{-1})_{\rm MSSM}^0-
\fr{33}{7\pi }n_{\Phi }\ln \fr{M_G}{M}-
\fr{33}{7\pi }n_{\Phi }P_{\de }^{(M)}+
\fr{39}{14\pi }P_{\de }^{(\mu_0)}~,
\la{alsKK4s}
\eeq
and its desirable value $0.119$ and successful unification is obtained for 
$n_{\Phi }=1$ and

$$
\l \de ~, ~\fr{M_G}{\mu_0}~,~\fr{M_G}{M}~,~M_G\r=
\l 1,~~~18.57,~~~10.56,~~~10^{3.13}~{\rm TeV}\r ~,
$$
\beq
\l 2,~~~4.78,~~~3.657,~~~10^{2.96}~{\rm TeV}\r ~,~
\l 3,~~~2.937,~~~2.465,~~~10^{2.92}~{\rm TeV}\r ~,\cdots
\la{sol4s}
\eeq
In this case $\eta $ must be zero, since its higher values drive $\al_a$
couplings
in non perturbative regime until they reach unification point.
For this scenario unification picture for $\de=1$, $\eta=0$ is plotted
on Fig. 1, (d).

As far the case with $5$ dimensional supermultiplets are concerned, also
for this scenario successful pictures of unification will be obtained for
$n_{\Phi }=1$, with presence of same $SU(3)_c$ adjoint (as in the case
above). Namely, $\al_s\simeq 0.119$ is obtained for

$$
\l \de ~, ~\fr{M_G}{\mu_0}~,~\fr{M_G}{M}~,~M_G\r=
\l 1,~~~18.02,~~~6.09,~~~10^{1.46}~{\rm TeV}\r ~,
$$
\beq
\l 2,~~~4.683,~~~2.742,~~~10^{1.39}~{\rm TeV}\r ~,~
\l 3,~~~2.895,~~~2.029,~~~10^{1.36}~{\rm TeV}\r ~,\cdots
\la{sol5s}
\eeq
Also now only $\eta=0$ case is allowed. Unification
picture for $\de=1$ is plotted on Fig. 1, (e).

For SUSY scenario with $6$-plets, successful unification take place
for $n_{\Phi }=1$ and 

$$
\l \de ~, ~\fr{M_G}{\mu_0}~,~\fr{M_G}{M}~,~M_G\r=
\l 1,~~~16.84,~~~3.9,~~~5.74~{\rm TeV}\r ~,
$$  
\beq
\l 2,~~~4.515,~~~2.169,~~~5.25~{\rm TeV}\r ~,~
\l 3,~~~2.823,~~~1.729,~~~5.12~{\rm TeV}\r ~,\cdots
\la{sol6s}
\eeq
Also, in this scenario only $\eta=0$ is allowed and 
unification picture for $\de=1$ is plotted on Fig. 1, (f).

As we have seen, successful unifications for non-SUSY and SUSY
scenarios can be obtained even for $\Phi $-plet masses M below the GUT
scale. For all this cases unification take place not too far from
the scale $M$ (for illustrations see Fig. 1). 

Through analyses, for neutrino masses in (\ref{numas4}), (\ref{numas5}), 
(\ref{numas6}) and (\ref{susynumas}), we have taken $M_f\simeq M_G$. 
However, it is possible to have unification scale, by few
factors and even more,  below the $M_f$. This would reduce scales $\mu_0$
and $M$, making scenarios easily testable on a future colliders.

\section{Conclusions}

In this paper we have suggested mechanism for natural suppressing
neutrino masses, within theories of low scale quantum gravity. Crucial
role is played by $\Phi $ scalars in different representations of
$SU(2)_L$. Selection of $\Phi $ is dictated from the value of fundamental
scale. Different scenarios were considered, in which neutrino masses are
suppressed in the needed level. Further studies, of low scale
theories
with those $\Phi $ states, would be an attempt to accommodate
atmospheric and solar neutrino anomalies \cite{atm, sol}. For this
purpose, one can also
introduce flavor symmetries and build different neutrino oscillation
scenarios in a spirit of \cite{osc}. The flavor symmetries within low
scale theories could play crucial role for suppression of FCNC together
with
natural understanding of hierarchies of the charged fermion masses and CKM
mixings \cite{flsym}. 
Particular interest deserve scenarios with fundamental scales (and
consequently
masses of $\Phi $ states) close to TeV range (cases with $5$ and
$6$-plets),being testable in a collider experiments of a nearest future.

Introduced $\Phi $ states (together with appropriate KK 
excitations) are also crucial for non-SUSY low scale unification, while for
SUSY scenarios $\Phi+\ov{\Phi } $ supermultiplets open up new
possibilities for successful unification. In considered examples, 
unification points are close to the fundamental scale (few or multi
TeV) and building realistic models, one have to take care for nucleon
stability. For latter, one of the mechanisms suggested in  
\cite{uni1, pdecay}
could be efficient. Detailed investigations and studies of these
and related issues will be presented elsewhere.

\bibliographystyle{unsrt}



\newpage

\vspace{-1cm}
\begin{figure}[tb]
\epsfysize=5.1in
\epsffile[125 195 500 500]{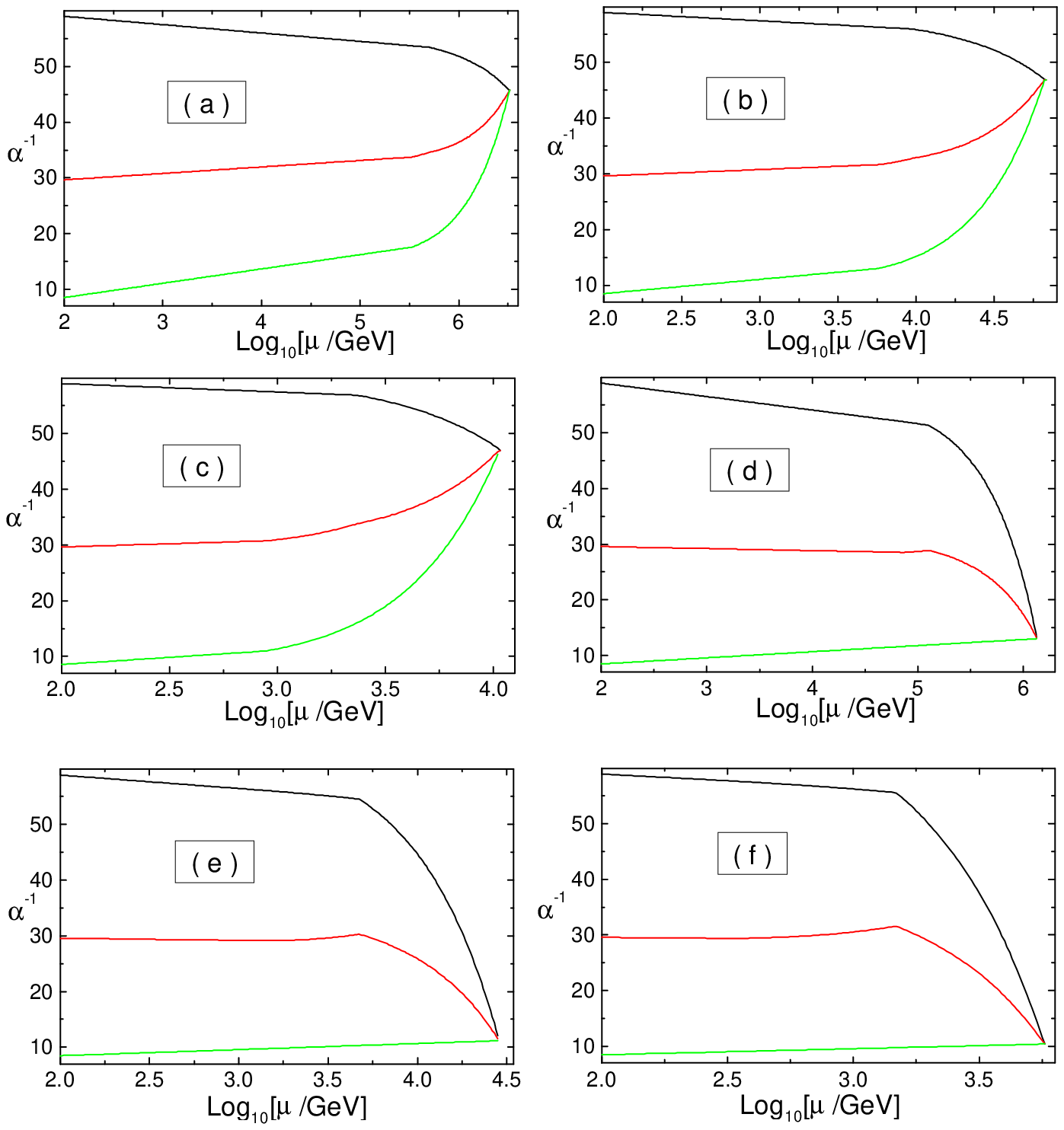} 
\begin{center}
\vspace{3cm} 
\caption[]{Unification pictures for $\al_s(M_Z)\simeq 0.119$
and $\de=1$, $\eta=0$;~
{\bf (a)}, {\bf (b)},  {\bf (c)} non-SUSY cases with two scalar 4, one 5
and one 6 plets respectively;~
{\bf (d)}, {\bf (e)}, {\bf (f)} SUSY cases with $\tan \bt\simeq 1$
and one pair of chiral 4, 5 and 6 supermultiplets respectively.
}
\end{center}
\end{figure}

\end{document}